\documentclass[review, 12pt]{elsarticle}

\usepackage{hyperref}
\journal{Carbon}

\usepackage{numcompress}
\usepackage{soul,xcolor}

\begin{document}

\begin{frontmatter}

\title{Band Structure Dependent Electronic Localization in Macroscopic Films of Single-Chirality Single-Wall Carbon Nanotubes}

% \tnotetext[mytitlenote]{Fully documented templates are available in the elsarticle package on \href{http://www.ctan.org/tex-archive/macros/latex/contrib/elsarticle}{CTAN}.}

% \address{Radarweg 29, Amsterdam}
% \fntext[myfootnote]{Since 1880.}

\author[utah]{Weilu Gao\corref{cor1}}%\fnref{fn1}}
\ead{weilu.gao@utah.edu, Tel:801-581-7054}
\cortext[cor1]{Corresponding author}
\address[utah]{Department of Electrical and Computer Engineering, University of Utah, Salt Lake City, Utah 84112, USA}

\author[buffalo]{Davoud Adinehloo}%\fnref{fn1}}
\address[buffalo]{Department of Electrical Engineering, University at Buffalo, Buffalo, NY 14228, USA}

\author[rice-ece]{Xinwei Li\fnref{cur-address}}
\address[rice-ece]{Department of Electrical and Computer Engineering, Rice University, Houston, TX 77005, USA}
\fntext[cur-address]{Present Address: Division of Physics, Mathematics and Astronomy, California Institute of Technology, Pasadena, CA 91125, USA}

\author[rice-ece]{Ali Mojibpour}
% \address[3]{Department of Electrical and Computer Engineering, Rice University, Houston, TX 77005, USA}

\author[tmu]{Yohei Yomogida}
\address[tmu]{Department of Physics, Tokyo Metropolitan University, Hachioji, Tokyo 192-0397, Japan}

\author[aist]{Atsushi Hirano}
\address[aist]{Nanomaterials Research Institute, National Institute of Advanced Industrial Science and Technology (AIST), Tsukuba, Ibaraki 305-8565, Japan}

\author[aist]{Takeshi Tanaka}
% \affiliation{Nanomaterials Research Institute, National Institute of Advanced Industrial Science and Technology (AIST), Tsukuba, Ibaraki 305-8565, Japan}

\author[aist]{Hiromichi Kataura}
% \affiliation{Nanomaterials Research Institute, National Institute of Advanced Industrial Science and Technology (AIST), Tsukuba, Ibaraki 305-8565, Japan}

\author[nist]{Ming Zheng}
\address[nist]{National Institute of Standards and Technology, Gaithersburg, MD 20899, USA}

\author[buffalo]{Vasili Perebeinos}
% \affiliation{Department of Electrical Engineering, University at Buffalo, Buffalo, NY 14228, USA}

\author[rice-ece,rice-phys,rice-mat]{Junichiro Kono}
% \address[rice-ece]{Department of Electrical and Computer Engineering, Rice University, Houston, TX 77005, USA}
\address[rice-phys]{Department of Physics and Astronomy, Rice University, Houston, TX 77005, USA}
\address[rice-mat]{Department of Materials Science and NanoEngineering, Rice University, Houston, TX 77005, USA}

% \author[mymainaddress,mysecondaryaddress]{Elsevier Inc}

% \author[mysecondaryaddress]{Global Customer Service\corref{mycorrespondingauthor}}
% \cortext[mycorrespondingauthor]{Corresponding author}

% \address[mymainaddress]{1600 John F Kennedy Boulevard, Philadelphia}
% \address[mysecondaryaddress]{360 Park Avenue South, New York}

\begin{abstract}
Significant understanding has been achieved over the last few decades regarding chirality-dependent properties of single-wall carbon nanotubes (SWCNTs), primarily through single-tube studies.  However, macroscopic manifestations of chirality dependence have been limited, especially in electronic transport, despite the fact that such distinct behaviors are needed for many applications of SWCNT-based devices.  In addition, developing reliable transport theory is challenging since a description of localization phenomena in an assembly of nanoobjects requires precise knowledge of disorder on multiple spatial scales, particularly if the ensemble is heterogeneous.  Here, we report an observation of pronounced chirality-dependent electronic localization in temperature and magnetic field dependent conductivity measurements on macroscopic films of single-chirality SWCNTs. The samples included large-gap semiconducting (6,5) and (10,3) films, narrow-gap semiconducting (7,4) and (8,5) films, and armchair metallic (6,6) films. Experimental data and theoretical calculations revealed Mott variable-range-hopping dominated transport in all samples, while localization lengths fall into three distinct categories depending on their band gaps. Armchair films have the largest localization length. Our detailed analyses on electronic transport properties of single-chirality SWCNT films provide significant new insight into electronic transport in ensembles of nanoobjects, offering foundations for designing and deploying macroscopic SWCNT solid-state devices.
  
\end{abstract}


\end{frontmatter}

\section{Introduction}

Since their discovery in the early 1990s, single-wall carbon nanotubes (SWCNTs) have served as an ideal nanoscale laboratory for investigating fundamental electronic, optical, magnetic, and thermal processes in one-dimensional (1D) systems~\cite{DresselhausetAl01Book,JorioetAl08Book}. Furthermore, macroscopic assemblies of SWCNTs, especially if they are ordered, are expected to lead to a wide variety of applications, such as lightweight electric wires for power transmission with ultrahigh current carrying capacity~\cite{YaoetAl00PRL,WangetAl14AFM} and mechanical strength~\cite{BehabtuetAl13Science,BaiEtAl2018NN}, ultrabroadband optoelectronic devices with strongly anisotropic optical constants~\cite{AndoEtAl1997JPSJ,RenetAl09NL,NanotetAl12AM,GaoetAl18NP}, thermal engineering via giant and anisotropic thermal conductivity~\cite{FujiiEtAl2005PRL,PopEtAl2006NL,YamaguchiEtAl2019APL}, and energy harvesting and conversion with substantial thermoelectric power~\cite{HicksEtAl1993PR,BlackburnEtAl2018AM,IchinoseEtAl2019NL}.

What determines most of the basic properties of a SWCNT is the chiral vector (or roll-up vector), defined as $\mathbf{C}_h = n\mathbf{a}_1 + m\mathbf{a}_2$, where $\mathbf{a}_1$ and $\mathbf{a}_2$ are the primitive lattice vectors of graphene~\cite{DresselhausetAl01Book}. Depending on the pair of integers ($n,m$), called the chirality indices, the SWCNT is metallic or semiconducting. Specifically, for $\nu \equiv (n-m)$ mod 3, nanotubes with $=\pm 1$ are semiconductors with large band gaps ($>$1\,eV), nanotubes with $\nu = 0$ and $n \ne m$ are semiconductors with curvature-induced small band gaps, and armchair nanotubes ($\nu=0$ and $n=m$) are ``metallic'' with electron-electron interaction-induced small band gaps.

Over the last two decades, considerable understanding has been achieved regarding ($n,m$)-dependent properties of SWCNTs through pioneering single-tube experiments~\cite{BockrathEtAl1999N,KimEtAl2001PRL,FuhrerEtAl2000S}. However, macroscopic manifestations of their promised extraordinary properties have been elusive because of defects, unintentional doping, intertube interactions, random orientations, and most significantly, mixed chiralities. When SWCNTs are synthesized in high-temperature furnaces, many chiralities of SWCNTs are produced together, including both semiconducting and metallic species, and as a consequence, exotic 1D charge carrier transport effects, such as quantum interference~\cite{KongEtAl2001PRL}, have never been observed in macroscopic SWCNT ensembles. Furthermore, chirality-dependent electronic transport studies have been challenging even in single-tube experiments, because of contact resistance issues that prevent studies of small-diameter SWCNTs ($< 1$\,nm)~\cite{BockrathEtAl1999N,DeshpandeEtAl2009S,SengerEtAl2018PR}. 

Recently, solution-based chirality separation techniques have been developed, offering opportunities for addressing challenges of structural polydispersity in SWCNT ensembles. %Current status of sorting methods mainly modify the hydrophilicity of SWCNTs in aqueous dispersion based on their electronic structure. 
Among these techniques are aqueous two phase extraction (ATPE)~\cite{KhripinEtAl2013JACS,SubbaiyanEtAl2014N,FaganEtAl2014AM} and gel chromatography~\cite{YomogidaEtAl2016NC} -- two methods widely used for large-scale separation. Especially combined with DNA-assisted selectivity~\cite{AoEtAl2016JACS}, ultrahigh-purity and single-chirality SWCNT suspensions can be prepared in substantial amounts. Furthermore, solution-based large-scale assembling methods, such as vacuum filtration~\cite{HeEtAl2016NN}, can preserve the chirality purity and produce wafer-scale uniform samples. %These two aspects of the advancement in materials preparation facilitate macroscopic metrology, including both electronic and optical methods. 
Transport studies have been performed on metal-enriched SWCNT films~\cite{YanagiEtAl2010N,WangEtAl2018PRM}, but multiple chiralities coexisting in the films made clear and comprehensive understanding difficult. Therefore, it is crucial to study transport in single-chirality SWCNT films, which has not been reported before.

In the present work, we systematically performed temperature and magnetic field dependent conductivity measurements of single-chirality SWCNT films, including semiconducting (6,5) and (10,3) films, chiral-metallic (7,4) and (8,5) films, and an armchair (6,6) film, prepared through solution-based chirality separation combined with a large-scale vacuum filtration technique. We found that Mott's variable-range-hopping (VRH) conduction mechanism dominates the transport behavior of both semiconducting and metallic SWCNT films at low temperatures. However, the metallic SWCNT film displayed significant deviation from what is expected from VRH at high temperatures, especially for armchair SWCNT films.
% We found that Mott's variable-range-hopping (VRH) conduction mechanism dominates the transport behavior of the semiconducting and chiral-metallic SWCNT films. However, the armchair SWCNT film displayed significant deviation from what is expected from VRH. 
Moreover, we extracted the localization lengths of carriers in these films, which strongly depended on the SWCNT band structure type, leading to the following general conclusions: large-gap semiconducting films with $\nu = \pm 1$ [e.g., (6,5) and (10,3)] have the smallest localization lengths, armchair films with $n=m$ [e.g., (6,6)] have the largest localization lengths, and narrow-gap semiconducting films with $\nu = 0$ and $n\neq m$ [e.g., (7,4) and (8,5)] have intermediate values. Note, there is an order of magnitude difference in localization lengths for chiral-metallic and armchair SWCNTs. Magnetoconductivity measurements on these films further corroborated these conclusions by providing the orders of magnitude for the localization lengths. Our theoretical calculations reproduced these trends and quantitatively explained the main features of the experimental data. Despite similar defect densities in these films, the armchair film exhibited stronger resilience against localization and also displayed significant deviation of calculated localization length compared to that extracted from experiments, indicating additional distinctly different transport mechanism from the other films.  These measurements and analyses thus provide significant new insight and guidance for future electronic and optoelectronic devices based on macroscopic assemblies of SWCNTs. 

\section{Experimental section}

\begin{figure}
  \includegraphics[width=0.9\textwidth]{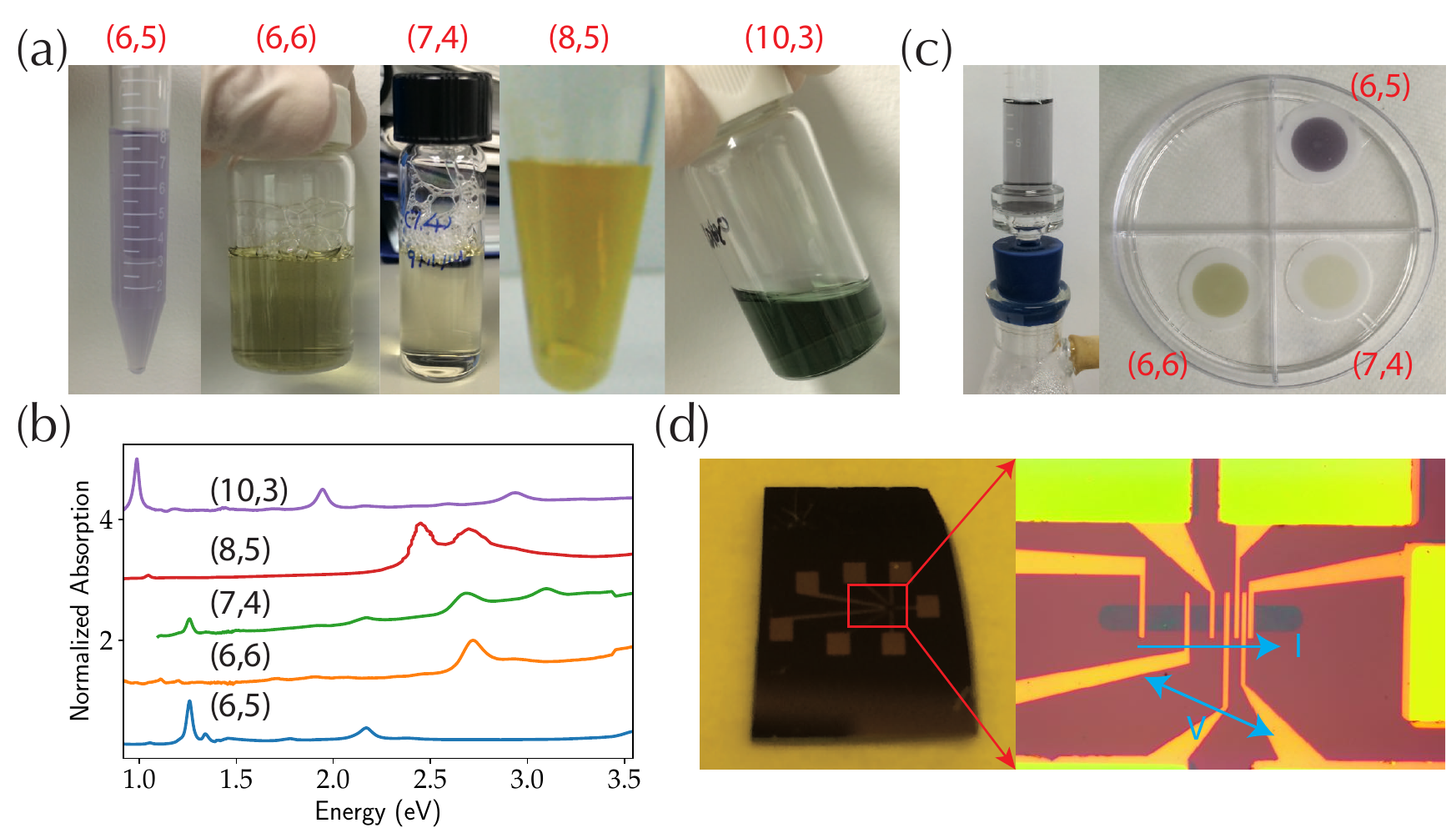}
  \caption{\label{mat_dev} {\bf Single-chirality SWCNT suspensions, films, and electronic devices.} (a)~Sorted single-chirality SWCNT suspensions of (6,5), (6,6), (7,4), (8,5), and (10,3). (b)~Optical absorption spectra for the sorted SWCNT suspensions. (c)~Large-scale uniform films of single-chirality SWCNTs obtained by using vacuum filtration. (d)~Fabricated devices for electronic transport measurements using standard micro/nanofabrication processes.}
\end{figure}

Figure\,\ref{mat_dev} summarizes the preparation methods and characteristics of the chirality-sorted suspensions, films, and fabricated electronic devices that we studied in this work. In preparing single-chirality SWCNT suspensions, we used two solution-based separation techniques: ATPE and gel chromatography.  Figure\,\ref{mat_dev}a is a photograph of the prepared suspensions of (6,5), (6,6), (7,4), (8.5), and (10,3) SWCNTs. (6,5), (6,6), and (7,4) suspensions were prepared using the standard ATPE method.\cite{KhripinEtAl2013JACS,SubbaiyanEtAl2014N,FaganEtAl2014AM} %Briefly, well-dispersed SWCNT suspensions consisting of multiple chiralities were mixed with a polymer solution including two polymers of different hydrophilicity. By carefully manipulating surfactant concentration and combination, SWCNTs of different band structures have different hydrophilicity and thus partition across two polymers in different fraction. By extracting one phase containing more shares of targeted species and adding a fresh new polymer phase, the partition and sorting process continue to purify suspension till nearly single-chirality suspensions. 
The (8,5) suspension was sorted and separated from an as-grown SWCNT powder following similar procedures, but instead of dispersing SWCNTs with multiple surfactants, the dispersant used was DNA~\cite{ZhengEtAl2003S,TuEtAl2009N,AoEtAl2016JACS,ZhengEtAl2017TCC}.
Finally, the (10,3) suspension was prepared using column gel chromatography~\cite{YomogidaEtAl2016NC}. %Briefly, well-dispersed SWCNT suspensions were fed into a conventional chromatography system, whose column was filled with gel beads. SWCNTs with different band structures have different bound strength with gel beads, where metallic and narrow-bandgap tubes are unbounded on beads. After elution of unbound SWCNTs with an aqueous surfactant solution, the adsorbed SWCNTs were eluted and collected through stepwise elution chromatography with gradually increased surfactant concentration. At certain concentration, single-chirality (10,3) SWCNTs were eluted. 
For the (6,5), (7,4), and (8,5) samples, the raw materials we used are CoMoCAT SG65i as-grown SWCNT powders (MilliporeSigma); for the (6,6) sample, the raw material we used is CoMoCAT CG200 as-grown SWCNT powder (MilliporeSigma); and for the (10,3) sample, the raw material we used is from HiPco SWCNTs (R1831, NanoIntegris). All sorted suspensions were inspected using absorption spectroscopy for evaluating the obtained purity. Figure\,\ref{mat_dev}b shows the obtained spectra, exhibiting signature excitonic peaks. Except the small amount of residual (6,5) nanotubes in the (7,4) suspension, all suspensions demonstrate high purities ($>90$\%). See Supporting Information Section 1 for suspension preparation details. 

We produced large-area films of single-chirality SWCNTs using vacuum filtration. Figure\,\ref{mat_dev}c shows the filtration system, together with photographs of the produced (6,5), (6,6), and (7,4) films. %The obtained suspensions sorted using aforementioned techniques were poured into the funnel of the filtration system, where a 80-nm pore size filter membrane was placed below the funnel to retain SWCNTs on top of it with smaller molecules, such as water and surfactants, penetrating through the membrane. As a result, large-scale SWCNT films have been deposited on the filter membrane. 
Due to the self-limiting nature of the vacuum filtration process~\cite{WuEtAl2004S,HeEtAl2016NN}, the film thickness was uniform and carefully controlled. Specifically, we tried to have the same film thickness used in this study, which was $\sim10\,$nm. Depending on the conditions used during the filtration process and colloidal properties of SWCNT suspensions, the controlled vacuum filtration method can produce aligned or randomly oriented SCWNT films.  The films used for the current transport study were randomly oriented; see Supporting Information Section 2, Fig.\,S2, and Fig.\,S3 for more information on film fabrication.
The obtained films were transferred onto nearly arbitrary substrates and are compatible with micro/nanofabrication processes. As shown in Fig.\,\ref{mat_dev}d, the films were first transferred onto silicon oxide/silicon substrates and patterned into device geometries using photolithography. We then annealed fabricated devices in an Ar atmosphere at 350\,$^{\circ}$C for 30\,min in order to minimize any effects from residues~\cite{HeEtAl2016NN}; see Supporting Information Section 2 and Fig.\,S2 for additional scanning electron microscopy and X-ray spectroscopy characterizations on obtained films.
%The outer large electrodes of the area $\sim$mm$^2$ in fabricated devices were bonded to a gold wire through indium cold welding, and the gold wire connects to electric connection in a cryostat. The indium cold welding helps protect devices from soldering damage.
The sample was mounted in a cryostat, and the temperature was varied from $2$\,K to $260$\,K. Four-point measurements were conducted to exclude the influence from contacts, as shown in Fig.\,\ref{mat_dev}d. The current was supplied from the two outmost electrodes, and the voltage was measured across a region of $5\,\mu$m in width and $8\,\mu$m in length. The magnetic field is applied perpendicular to the device plane. At each temperature and magnetic field, a current-voltage sweep was performed, and the conductivity value was extracted in the linear region. 

\section{Results and discussion}

\begin{figure}
  \includegraphics[width=0.6\textwidth]{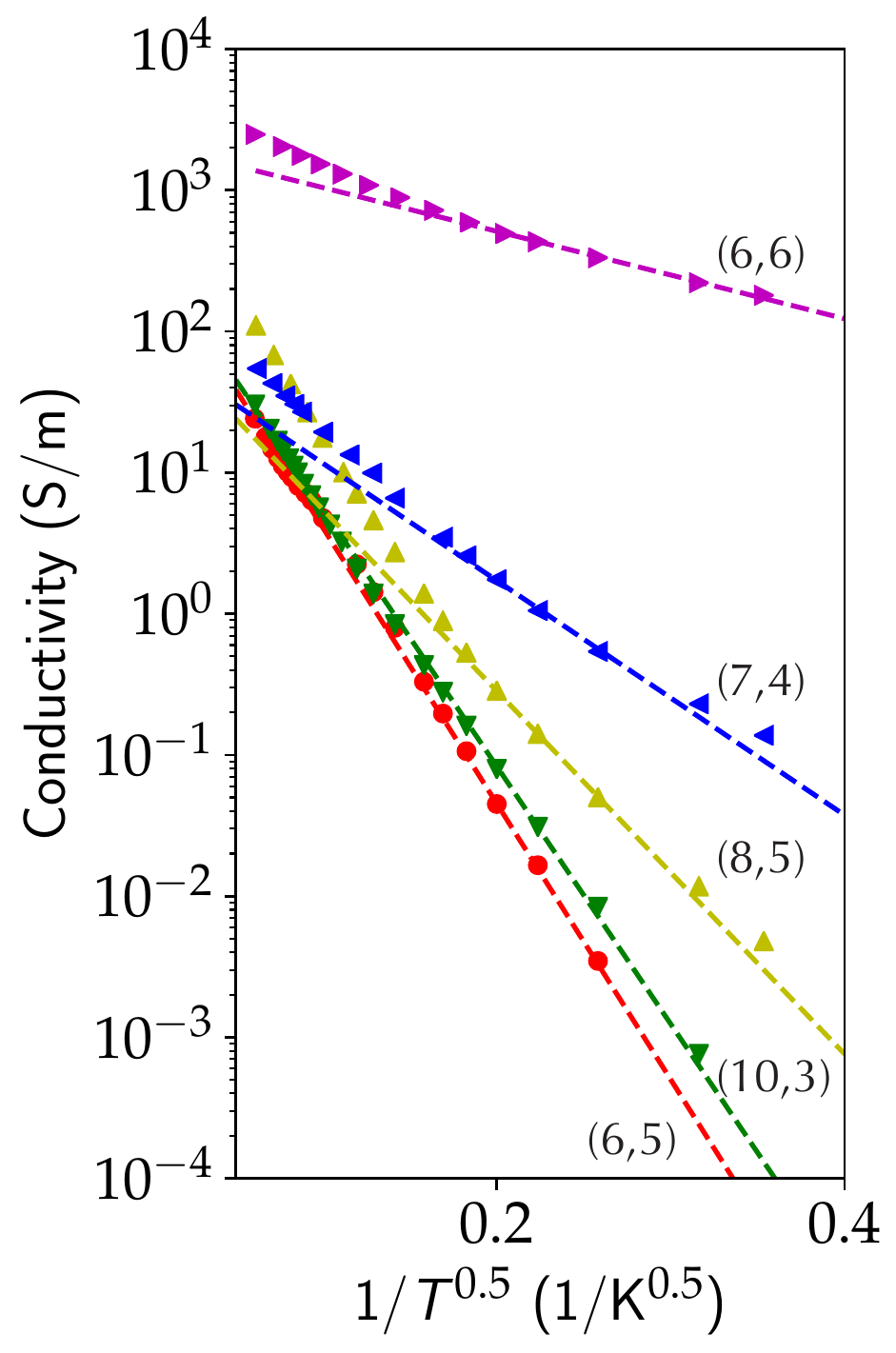}
  \caption{\label{cond} {\bf Temperature-dependent electrical conductivity of the five single-chirality SWCNT films.} Solid markers of different styles are experimental data, and dashed lines are corresponding fitting curves based on the 1D Mott variable range hopping model.}
\end{figure}

Figure\,\ref{cond} shows the conductivity ($\sigma$) versus temperature ($T$) for all five films. They all show a general trend of decreasing conductivity with decreasing temperature. Our analysis suggests that electron-phonon and electron-electron interactions for either intratube transport or intertube transport cannot alone account for such strongly temperature-dependent conductivity, but can have a significant contribution for specific types of SWCNTs under certain conditions. As a result, all temperature-dependent conductivity data were fit with the 1D Mott VRH model, $\sigma(T) = \sigma_0\textrm{exp}\left [ -(\frac{T_0}{T})^{1/2} \right ]$, where both $\sigma_0$ and $T_0$ are constants to be determined via fitting. Note that the Efros-Shklovskii (ES) VRH model has the same expression with strong temperature dependence, and the SWCNT density in films determines which model is more applicable. Specifically, the factor $n_\textrm{CNT}L_\textrm{CNT}^3$, where $n_\textrm{CNT}$ is the volume density of nanotubes in the film and $L_\textrm{CNT}$ is the length of SWCNTs, is the decisive parameter as to which model is applicable, i.e., the Mott VRH model ($n_\textrm{CNT}L_\textrm{CNT}^3 > 1$) or the ES VRH model~\cite{HuEtAl2006PR} ($n_\textrm{CNT}L_\textrm{CNT}^3 < 1$). Because of the high packing density and random SWCNT orientations in the produced films (see Supporting Information Fig.\,S2), $n_\textrm{CNT}$ can be estimated to be $\approx 2/(dL^2_\textrm{CNT})$, where $d$ is the SWCNT diameter, and thus, $n_\textrm{CNT}L_\textrm{CNT}^3 \approx 2L_\textrm{CNT}/d$.  The aspect ratio of SWCNTs in our produced films is around 200--300, which suggests that $n_\textrm{CNT}L_\textrm{CNT}^3\gg 1$, i.e., the Mott VRH model is more appropriate for our system~\cite{HuEtAl2006PR}; see Supporting Information Fig.\,S4 for atomic force microscopy characterization of SWCNT length.

\begin{table}
  \centering
  \begin{tabular}{ |c|c|c|c|c|c| } 
   \hline
    SWCNTs & (6,5) & (10,3) & (8,5) & (7,4) & (6,6) \\ 
   \hline
    Exp. $T_0$\,(K) & $2.0\times10^3$ & $1.8\times10^3$ & $3.7\times10^2$ & $8.8\times10^2$ & $5.1\times10^1$ \\ 
   \hline 
    $E_\textrm{g}$\,(eV) & 1.27 & 0.99 & 0.446 & 0.379 & 0.193 \\ 
    \hline
    Cal. $T_0$\,(K) & $2.3\times10^3$ & $1.4\times10^3$ & $5.8\times10^2$ & $8.0\times10^2$ & $3.6\times10^2$ \\ 
   \hline
    Cal. $\xi$\,(nm) & 0.21 & 0.44 & 2.7 & 1.7 & 8.0 \\ 
   \hline
  \end{tabular}
  \caption{Characteristic temperature and localization length of the five single-chirality SWCNT films determined by experiments and calculations. Note that electron correlation and curvature induced gaps are added to nominally metallic SWCNTs in the last 3 columns. }
  \label{table:1}
\end{table}

The obtained values of $T_0$ for the five films fall into three distinct categories (see the Exp.\,$T_0$ row of Table 1).  The large-gap semiconducting SWNCT films [(6,5) and (10,3)] have $T_0$ on the order of $10^3$, the narrow-gap metallic SWCNT films [(8,5) and (7,4)] have $T_0$ on the order of $10^2$, and the armchair SWCNT (6,6) film has $T_0$ on the order of $10^1$. The localization length $\xi$, which corresponds to the localization radius of SWCNT states within single CNTs, is inversely proportional to $T_0$ through $T_0 = \frac{\beta}{\xi\cdot\textrm{DOS}(\varepsilon)}$, where $\textrm{DOS}(\varepsilon)$ is the density of states at energy $\varepsilon$ and $\beta$ is a constant, which we take $\beta=1$ here. Because the five films were prepared using the same procedures and are thus expected to have similar defect densities, the existence of three distinct categories of $T_0$ and $\xi$ values suggests that the localization mechanism is strongly band-structure dependent. Note that the variations of quality in the raw SWCNT materials did not have major contributions to the defect densities of the studied films, as suggested by the fact that the Raman intensity ratio of the G peak to the D peak ($I_D/I_G$) was much larger in the raw materials ($I_D/I_G \sim 20$) than that in the prepared films ($I_D/I_G \sim 3.5$)~\cite{DresselhausEtAl2010PTRS}; see Supporting Information Fig.\,S5.
% Furthermore, the (6,6) film shows a markedly different behavior from what is predicted by the 1D Mott VRH model, which suggests an entirely different transport mechanism. 
Also, a comparison of these three categories shows that the (6,6) film displays the longest localization length. 

\begin{figure}
  \includegraphics[width=0.9\textwidth]{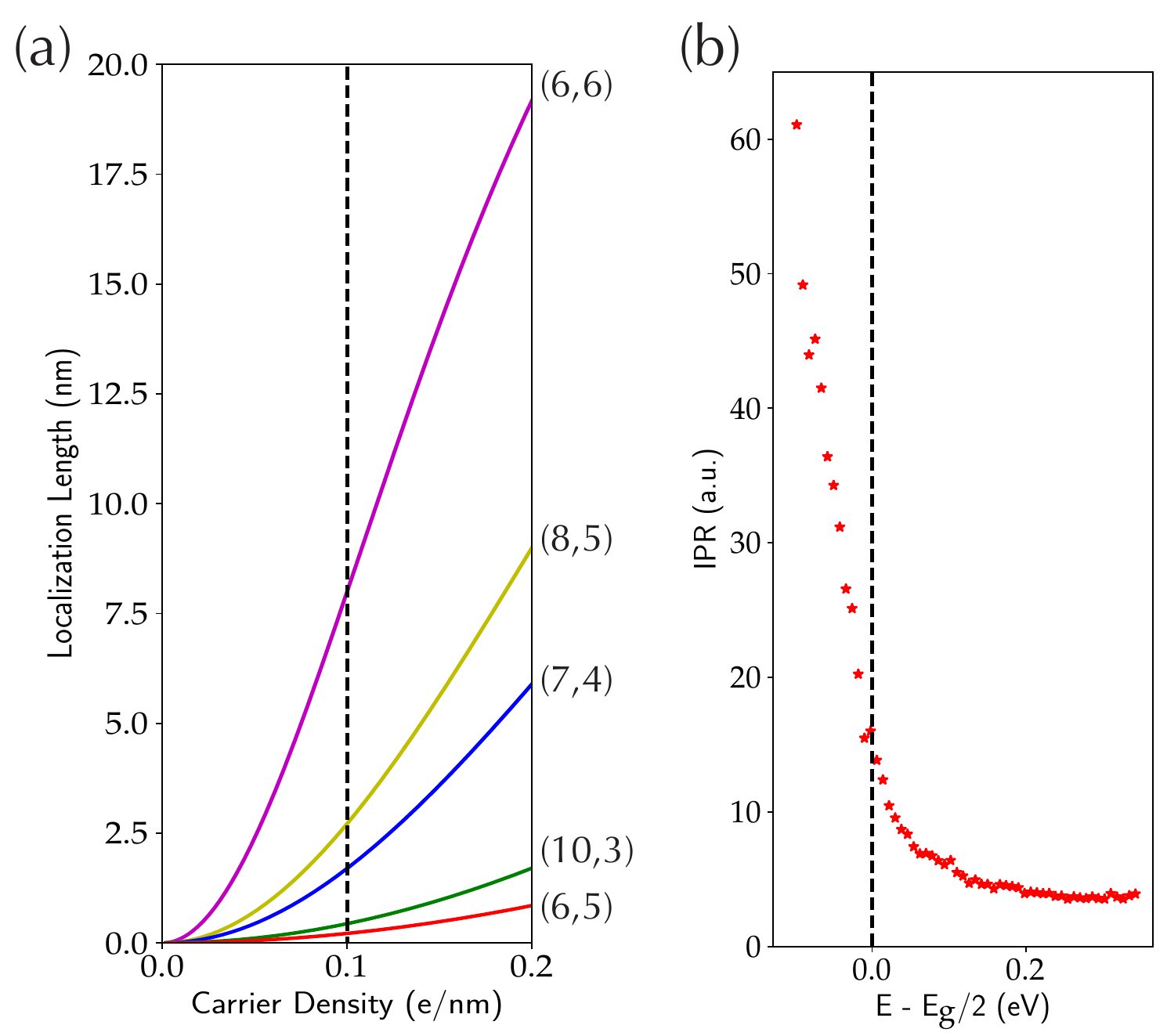}
  \caption{\label{xi} {\bf Calculated localization lengths  and inverse participation ratio in single-chirality SWCNT films.}  a) shows localization length using Eq.~(\ref{eq6}) in five films. The solid vertical line represents the assumed doping level for all samples. b) shows IPR using Eq.~(\ref{eq7}) in (6,5) SWCNT with the supercell length 200 nm. The large IPR values near the band edge, shown by the dashed vertical line, indicate carrier localization. }
\end{figure}

To better understand the relation between the band structure and localization mechanism, Fig.\,\ref{xi}a summarizes our calculated localization lengths as a function of doping density for five single-chirality SWCNT samples. The localization length $\xi$ in the presence of disorder can be estimated as follows:\cite{TakashimaEtAl2016JAP}
\begin{eqnarray}
\xi(\varepsilon)&=&\vert v(\varepsilon)\vert \tau(\varepsilon),
\label{eq1}
\end{eqnarray}
where $\varepsilon$ is the energy and $\tau$ is the scattering time due to the disorder, which can be evaluated using Fermi's golden rule
\begin{eqnarray}
\frac{1}{\tau(\varepsilon)}&=&\frac{2\pi}{\hbar}\langle\vert \langle \Psi_{\varepsilon} \vert {\rm H} \vert \Psi_{\varepsilon}\rangle \vert^2 \rangle_{\rm dis} {\rm DOS}(\varepsilon),
\label{eq2}
\end{eqnarray}
where ${\rm DOS}(\varepsilon)=2L/{\pi\hbar v(\varepsilon)}$ is the density of states of a SWCNT of length $L$ for a single spin channel. We assume that disorder scattering does not mix spin, but only valleys, as it is evident from the D band Raman signal (not shown). 

In the Anderson model with a random on-site potential disorder in the range $[-W/2,W/2]$, the averaged square of the matrix element is given by~\cite{NunezEtAl2016JAP}
\begin{eqnarray}
\langle\vert \langle \Psi_{\varepsilon} \vert {\rm H} \vert \Psi_{\varepsilon}\rangle \vert^2 \rangle_{\rm dis}&=&\frac{W^2}{12N},
\label{eq3}
\end{eqnarray}
where $N=L\pi d/A_\mathrm{c}$ is the number of carbon atoms in a SWCNT of diameter $d=a\sqrt{n^2+nm+m^2}/\pi$ and $A_\mathrm{c}=a^2\sqrt{3}/4$ is the area per carbon atom in graphene with a unit cell length $a=2.46$\,\AA.

Putting the results of Eqs.\,(\ref{eq1})--(\ref{eq3}) together, we obtain
\begin{eqnarray}
\xi(\varepsilon)&=&3\pi d \frac{\hbar^2v(\varepsilon)^2}{W^2A_\mathrm{c}}.
\label{eq4}
\end{eqnarray}
A hyperbolic dispersion velocity is given by $v(\varepsilon)=v_\mathrm{F}\sqrt{\varepsilon^2-E_\mathrm{g}^2/4}/\varepsilon$, where $E_\mathrm{g}$ is the SWCNT band gap and $v_\mathrm{F}=10^8$ cm/s is the Fermi velocity in graphene.
We assume that all SWCNTs have the same level of disorder $W$ since the fabrication and processing conditions were very similar. Thus, effects due to defects and disorder in the different samples are believed to be similar. In addition, since the samples were annealed before the transport measurements, unintentional doping from adsorbed molecules in the environment, such as water and oxygen, can reasonably be assumed to be the same. As a result, we also assume that the doping level $n$ is the same for all SWCNTs, which relates to the Fermi energy $\varepsilon_\mathrm{F}$ at zero temperature according to 
\begin{eqnarray}
  \varepsilon_\mathrm{F}=\sqrt{(E_\mathrm{g}/2)^2+(\pi\hbar v_\mathrm{F} n/4)^2}. 
  \label{eq5}
\end{eqnarray}
  
Therefore, to compare the values of $\xi$ in different SWCNTs, we use the following relation
\begin{eqnarray}
\xi(n)&=&3\pi d\frac{\hbar^2v_\mathrm{F}^2}{W^2A_\mathrm{c}}\frac{\left(\pi n\hbar v_F/(2E_g)\right)^2}{1+ \left(\pi n\hbar v_F/(2E_g)\right)^2}.
\label{eq6}
\end{eqnarray}
Simulations of $\xi$ according to Eq.\,(\ref{eq6}) for $W=1.9$\,eV are shown in Fig.\,\ref{xi}a, for the chosen values of the band gap given in Table\,\ref{table:1}. For the semiconducting SWCNTs, we use conventional values~\cite{WeismanEtAl2003NL}. For the metallic armchair (6,6) SWCNTs, we used a correlation band gap value suggested in Ref.\,\cite{DeshpandeEtAl2009S}. For the chiral metallic SWCNTs, we used a combination of the curvature induced gap~\cite{SengerEtAl2018PR} and the electron correlation induced gap. For a typical carrier density of $n=0.1$\,e/nm, shown by the vertical line in Fig.\,\ref{xi}a, we can find a good match between the simulated values of $T_0$ and the experimentally obtained values, except for the (6,6) film. In the latter case, we find a factor of 7 larger value of calculated $T_0$ than the experimental value. This indicates that the origin of temperature dependence is different in the (6,6) sample, and armchair SWCNTs are less affected by defects and thus weakly localized. Specifically, we found that intertube contributions to the total conductivity become more significant in armchair and metallic SWCNTs. A detailed discussion is beyond the scope of this paper and will be reported elsewhere. Indeed, under our assumption for the same values of disorder strength and carrier density in different SWCNTs, we find that the localization length in (6,6) SWCNTs is 8\,nm. Therefore, the condition for the applicability of the Mott VRH mechanism $L_\textrm{CNT}\gg\xi$ is weakened in the (6,6) film as compared to the other SWCNT samples, where $L_\textrm{CNT}\sim 200$\,nm. In a prior individual multiwall CNT transport experiment, a localization length of 4.5\,nm was reported~\cite{WangEtAl2007SSC}, which is close to the values we obtained for the (6,6) and (8,5) CNT films.

Furthermore, the degree of localization can be expressed by inverse participation ratio (IPR), which is defined as
\begin{eqnarray}
\textrm{IPR}(E_\mathbf{k})&=&N\frac{\sum_{i=1}^N|A_{i,\mathbf{k}}|^4}{(\sum_{i=1}^N|A_{i,\mathbf{k}}|^2)^2},
\label{eq7}
\end{eqnarray}
where $N$ is number of orbitals in the unit cell and $A_{i,\mathbf{k}}$ is $i$-th eigenstate of wavevector $\mathbf{k}$. A fully delocalized state corresponds to $\textrm{IPR}=1$, while a fully localized state on a single atom corresponds to $\textrm{IPR}=N$. Fig.\,\ref{xi}b depicts averaged IPR in (6,\,5) SWCNT for $W=1.9$ eV. The lower energy states near the band edge are localized, especially those falling into the bandgap. For higher energies, the states tend to be less localized.

\begin{figure}
  \includegraphics[width=0.5\textwidth]{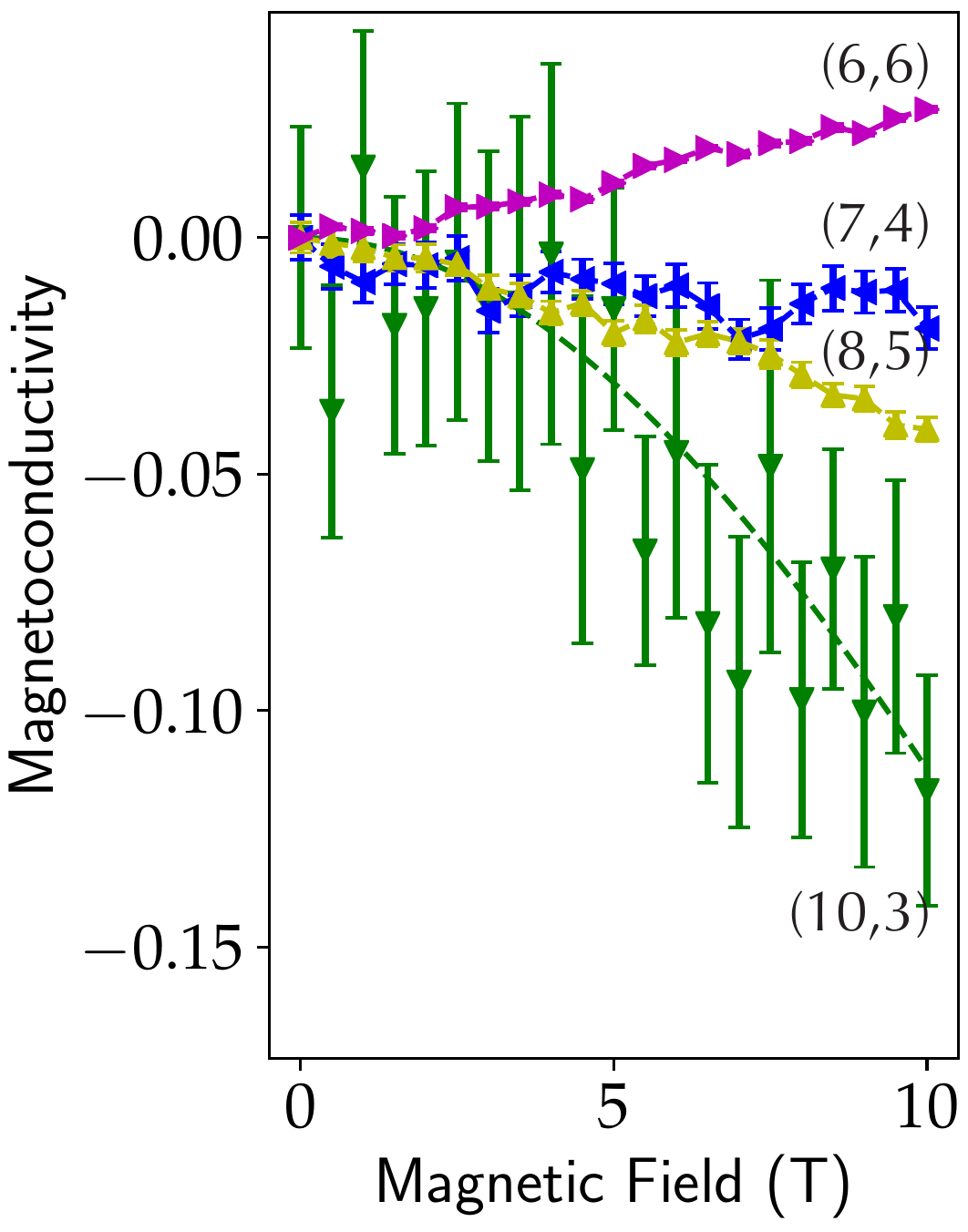}
  \caption{\label{mr} {\bf Magnetoconductivity of the single-chirality SWCNT films.} Solid dots are experimental data for four films. The (6,5) film had a large resistance value out of our measurement range. Green dashed line is the fitting curve for the (10,3) film, while the lines for the others simply connecting the experimental data points.}
\end{figure}

Finally, we performed magnetoconductivity (MC) measurements on the five films at $T=20$\,K and three distinct kinds of behaviors were again clearly observed; see Fig.\,\ref{mr}. Specifically, the (10,3) film displayed strong negative MC while the (6,6) film showed positive MC, up to a magnetic field ($B$) of 10\,T. The (7,4) and (8,5) films instead demonstrated a combination of the two trends. In the (10,3) film, strong localization paradigms including two negative MC contributions created by wave-function shrinkage and spin-dependent hopping feature a scaling dependence of $B^2$. We fit the experimental data (green dots in Fig.\,\ref{mr}) with the expression ln$[\rho(B)/\rho(0)] = \frac{2}{3}\frac{B^2}{B_0^2}$, where $B_0 = \frac{\sqrt{2}\phi_0}{\pi \xi^2 \sqrt{\frac{T_0}{T}}}$, $\phi_0 = \frac{h}{e}$ is the flux quantum, and $\rho(B)$ is the resistance in magnetic field $B$. The extracted $B_0$ is $\sim23$\,T and $\xi$ for the (10,3) film is $\sim1.4$\,nm. This value qualitatively agrees with the value obtained from theoretical calculations based on temperature-dependent conductivity measurements. Moreover, the (6,6) film displayed a signature of weak localization and positive MC, suggesting the largest localization length. Finally, the behaviors of the (7,4) and (8,5) films stayed in the middle with medium localization strengths. The orders of magnitude and trends of the localization strengths obtained from MC measurements are consistent with results from temperature-dependent conductivity. 

\section{Conclusions}

In summary, we measured temperature- and magnetic field-dependent conductivity of single-chirality SWCNT films with known chiralities ($n$,$m$) of three categories: (A)~$\nu = (n-m)$ mod 3 $=\pm 1$, (B)~$\nu = 0$ and $n \ne m$, and (C)~$\nu=0$ and $n=m$ (armchair). Despite similar defect densities in the films, the obtained localization length exhibited distinctly ($n$,$m$)-dependent behaviors: Films in category (A) have the smallest localization length, films in category (B) have medium localization lengths, and films in category (C) have the largest localization length. Our theoretical calculations based on Mott VRH explained all observations, except for the armchair SWCNT film (category C). Specifically, the VRH formula breaks down for armchair SWCNTs and an additional mechanism might be responsible for the observed temperature dependence. The largest localization lengths in macroscopic samples of armchair SWCNTs make them promising for future electronic and optoelectronic applications.

\pagebreak
% \begin{acknowledgement}
\noindent\textbf{Acknowledgement} -- J.\,K. acknowledges the support from the Department of Energy Basic Energy Sciences through grant no.\ DEFG02-06ER46308 (optical spectroscopy experiments), the Robert A.\ Welch Foundation through grant no.\ C-1509 (sample preparation), and the support from the JST CREST program, Japan, through Grant Number JPMJCR17I5.  W.\,G. thanks the support from the University of Utah start-up fund. V.\,P. acknowledges support from the Vice President for Research and Economic Development (VPRED) and the Center for Computational Research at the University at Buffalo (\url{http://hdl.handle.net/10477/79221}). 
%   ~\footnote{Center for Computational Research, University at Buffalo, http://hdl.handle.net/10477/79221.}.
%   \item[Competing Interests] The authors declare that they have no competing financial interests.
%   \item[Correspondence] Correspondence and requests for materials should be addressed to Weilu Gao (email: weilu.gao@utah.edu).
% \end{acknowledgement}

% \begin{suppinfo}

% \noindent The Supporting Information is available free of charge.
% Information on the details of sample preparation and characterization (PDF).
% \end{suppinfo}

%% With normal representation of data
%  \pagebreak
%  \begin{table}
%   \centering
%   \begin{tabular}{ |c|c|c|c|c|c| } 
%   \hline
%     CNTs & (6,5) & (10,3) & (8,5) & (7,4) & (6,6) \\ 
%   \hline
%     Exp. $T_0$\,(K) & 2025 & 1765 & 366 & 876 & 51 \\ 
%   \hline 
%     $E_\textrm{g}$\,(eV) & 1.27 & 0.99 & 0.446 & 0.379 & 0.193 \\ 
%     \hline
%     Cal. $T_0$\,(K) & 2250 & 1420 & 580 & 800 & 355 \\ 
%   \hline
%     Cal. $\xi$\,(nm) & 0.21 & 0.44 & 2.7 & 1.7 & 8.0 \\ 
%   \hline
%   \end{tabular}
%   \caption{Characteristic temperature and localization length of the five single-chirality SWCNT films determined by experiments and calculations}
%   \label{table:1}
% \end{table}

%% With scientific representation of data
 \pagebreak

\bibliography{weilu,jun}

\end{document}